\documentstyle[aps,prb,preprint,epsf]{revtex}

\begin{document}
\draft
\widetext
\preprint{Preprint: IUCM97-017 and 
          DFM-97-R13}
\title{Frictional Coulomb drag in strong magnetic fields}

\author{Martin Christian B\o nsager,$^{1}$ 
Karsten Flensberg,$^{2}$
Ben Yu-Kuang Hu,$^3$ 
and Antti-Pekka Jauho$^3$}
\address{$\mbox{}^1$ Department of Physics, 
Indiana University, Bloomington, Indiana 47405-4202}
\address{$\mbox{}^2$ Danish Institute of Fundamental Metrology,
Bldg. 307, Anker Engelunds Vej 1,
DK-2800 Lyngby, Denmark}
\address{$\mbox{}^3$ Mikroelektronik Centret,
Bldg. 345east, Technical University of Denmark,
DK-2800 Lyngby, Denmark}
\date{\today}
\maketitle

\begin{abstract}
A treatment of frictional Coulomb drag between two 2-dimensional 
electron layers in a strong perpendicular magnetic field,
within the independent electron picture, is presented. 
Assuming fully resolved Landau levels, the linear response theory 
expression for the transresistivity $\rho_{21}$ is evaluated using 
diagrammatic techniques. The transresistivity is given by an integral 
over energy and momentum transfer weighted by the product of the 
screened interlayer interaction and the phase-space for scattering 
events. We demonstrate, by a numerical analysis of the transresistivity, 
that for well-resolved Landau levels the interplay between these two 
factors leads to characteristic features in both the magnetic field- and 
the temperature dependence of $\rho_{21}$. Numerical results are compared 
with recent experiments.
\end{abstract}


\section{Introduction}\label{intro}
When two 2-dimensional charged systems are placed in close proximity, 
transport in one
layer will drive the adjacent layer out of equilibrium. Even
if the barrier separating the two layers is high and wide enough
to prevent tunneling, interlayer interactions can still be sufficiently
strong that a current drawn in one layer can {\it drag} along
a current in the other layer. This phenomenon was theoretically 
proposed by Pogrebinskii\cite{pogrebinskii} and by Price,
\cite{price} and has become known as {\it frictional drag}. In most
frictional drag experiments a current ${\bf J}_1$ is drawn in one
layer; the second layer is an open circuit and no current is 
allowed to flow. To oppose the dragging force, an electric field
${\bf E}_2$ develops in the second layer. The ratio of ${\bf E}_2$ 
and ${\bf J}_1$ is called the {\it transresistivity} $\rho_{21}$
[see Eq.~(\ref{rho21def}) below] and is a measure of a rate of momentum 
transfer from the first to the second layer. 

Experimental realizations of frictional drag between 2--dimensional systems 
were first reported by Gramila {\it et al}. for two electron 
layers,\cite{gramila} and by Sivan {\it et al}. for electron--hole
systems.\cite{sivan}
These experiments inspired a large number of theoretical works, 
and the experiments (which were all done in zero magnetic field) 
are by now fairly well 
understood.\cite{elechole,anttihenrik,zheng,plasmonlett,plasmonlong,vignale,plasmonhill,phonon}

Recently, attention has turned towards frictional drag in 
the presence of a magnetic field --- the topic of the present 
paper. Experiments of drag in a magnetic field have 
been reported by Hill {\it et al}.,\cite{hill} Rubel 
{\it et al}.,\cite{rubel} Feng {\it et al.},\cite{feng} and
Eisenstein {\it et al.}\cite{eisenstein}

Frictional drag is of fundamental interest because it can 
serve as a sensitive probe of two important aspects of transport 
in mesoscopic systems, namely the screened interlayer interaction 
and the form of the irreducible polarization function $\chi(q,\omega)$,
which is central to many theoretical considerations. 
In the presence of a magnetic field, in particular, the screening 
of the interaction and the polarization function can assume 
different forms depending on the number of filled Landau levels 
and the degree of disorder. 

Over the past two decades new and often surprising aspects of
the physics of 2-dimensional electron gases (2DEGs) in  magnetic fields 
have continued to emerge.\cite{qhe} 
Not only single layer systems show intriguing physics; 
double-layer quantum Hall systems also exhibit a number of 
interesting aspects,\cite{dlqhe} and frictional drag is 
expected to do the same. 

In this paper we present a treatment of Frictional Coulomb drag in 
strong magnetic fields where the Landau levels are fully resolved, 
{\it i.e.}, $\omega_c\tau\gg 1$ where $\omega_c=eB/m^*$ is the cyclotron 
frequency and $\tau$ is the transport scattering time. 
Some numerical aspects of our work have been presented in a previous 
publication,\cite{prl}
here we give the full details of the analytic background underlying
these results, extend them to other parameter values, and
compare them critically with recent experiments.\cite{hill,rubel}

We work under the assumption that an independent electron picture 
applies. At sufficiently low temperatures and/or high magnetic fields 
this condition will not be satisfied and additional physics must 
be included as the following two examples reveal. 

In disordered systems localization becomes important at low temperatures. 
Chalker and Daniell \cite{chalker} found 
that the diffusion of electrons in the lowest Landau level is 
anomalous, {\it i.e.} the 
`diffusion constant' scales as $D(q,\omega)=D_0(\omega/q^2)^{\eta/2}$,
where $\eta$ is found numerically to be $\eta\simeq 0.38$. 
Shimshoni and Sondhi pointed out that frictional drag would be 
a way to experimentally measure $\eta$, since the drag effect 
in that case would be proportional to $T^{2-\eta}$ at the lowest
temperatures.\cite{shimshoni} Another example consists of
high mobility systems in high magnetic fields,
where intralayer electron--electron interactions are important at
low temperatures. At filling factor $\nu = 1/2$ the 2DEG can be 
discussed in terms of composite fermions. The polarization function 
$\chi(q,\omega)$ assumes a unique form which was first derived by 
Halperin, Lee and Read.\cite{hlr} Three recent papers\cite{gaugefluc} 
have considered frictional drag in this regime and shown that the 
transresistivity should be proportional to $T^{4/3}$ as the 
temperature $T$ approaches zero.

The outline of the paper is as follows. 
In Section~\ref{forma} we define the model of the system and establish
the theoretical framework. The transresistivity is in general given
in terms of three-body correlation functions which are considered in
Section~\ref{triangle}. In Section~\ref{tribub} we examine in which limits 
the three-body correlation functions are proportional to the 
imaginary part of the polarization function. This relation has been
tacitly assumed by most other authors. 
We discuss the result and the relation to similar results in zero 
magnetic field and a brief discussion of Hall-drag is provided.
Numerical evaluations are presented in Section~\ref{num}; we focus on 
the dependence of the transresistivity upon magnetic field strength 
and temperature. Section~\ref{conclu} summarizes the conclusions.

\section{Formalism}\label{forma}
We consider a system of two 2-dimensional electron gases 
separated by a distance $d$. A uniform, constant
magnetic field ${\mathbf B}=B{\mathbf \hat{z}}$ is applied 
perpendicular to the two layers which define the $xy$-plane. 
The two layers have electron densities
of $n_1$ and $n_2$, respectively.  
When a current density ${\mathbf J}_1$ is drawn in layer 1, the 
interlayer interactions will induce an electric field ${\mathbf E}_2$
in layer 2, which is an open circuit ({\it i.e.,} no current is allowed to
flow in layer 2).
The induced electric field can be measured by a voltage probe, and
the {\it transresistivity} tensor is defined according to
\begin{equation}\label{rho21def}
{\mathbf E}_2=\tensor{\rho}_{21}{\mathbf J}_1.
\end{equation}
The transresistivity is what is measured experimentally and hence 
the object to determine. However, linear response theory, 
on which our theoretical approach is based, yields the
trans{\it conductivity} $\tensor{\sigma}_{21}$, defined 
by an experiment where the first layer is biased with an
electric field and the induced current density 
is measured in the second layer:
${\mathbf J}_2=\tensor{\sigma}_{21}{\mathbf E}_1$. 
The transresistivity can be obtained from the transconductivity by
\begin{eqnarray}\label{basic}
\tensor{\rho}_{21}&=&
\left[-\tensor{\sigma}_{11}\tensor{\sigma}_{21}^{-1}
\tensor{\sigma}_{22} + \tensor{\sigma}_{12}\right]^{-1}\nonumber\\
&\simeq& -\tensor{\rho}_{22}
\tensor{\sigma}_{21}\tensor{\rho}_{11},
\end{eqnarray}
where 
the approximate equality is valid because
we assume the magnitude of the individual layer conductivities, 
$(\tensor{\rho}_{11})^{-1}$ and $(\tensor{\rho}_{22})^{-1}$,
to be much larger than the transconductivity. 

To calculate the transconductivity we follow the general framework 
developed independently by Kamenev and Oreg,\cite{kamenev} and by 
Flensberg, Hu, Jauho and Kinaret.\cite{bakk}
The transconductivity is calculated using the Kubo formula
for linear response,\cite{mahan} {\it i.e.}, it is expressed as a 
current--current 
correlation function
\begin{equation}
\sigma_{21}^{\alpha\gamma}({\mathbf k},\Omega)=\frac{ie^2}{\hbar\Omega}
\Pi_{21}^{\alpha\gamma,r}({\mathbf k},\Omega),
\end{equation}
where $\Pi_{21}^{\alpha\gamma,r}({\mathbf k},\Omega)$ is the Fourier 
transform of the retarded current--current correlation function
\begin{equation}
\Pi_{21}^{\alpha\gamma,r}({\mathbf x}-{\mathbf x'},t-t')
=-i\theta(t-t')
\langle [j_2^{\alpha}({\mathbf x},t)\ ,\ j_1^{\gamma}({\mathbf x'},t')]
\rangle.
\end{equation}
Here ${\bf j}_1$ and ${\bf j}_2$ are (kinematic) particle current 
operators in the two different layers
(denoted by the subscripts 1 and 2), and $\alpha$ and $\gamma$
are Cartesian coordinates of the two-by-two transconductivity tensor. 
$\langle \cdots \rangle$ is a statistical average and $[\cdots ,\cdots]$ is
a commutator. The position vectors ${\mathbf x}$ and ${\mathbf x'}$ 
reside in the 2-dimensional planes of the 2DEGs. 
In this paper, we assume like charges in the two layers; the sign
of $\sigma_{21}$ is reversed for unlike charges.

To lowest order in the screened interlayer interaction, the
transconductivity can be expressed as\cite{kamenev,bakk}
\begin{eqnarray}
\sigma_{21}^{\alpha\gamma}&=&\frac{e^2}{2\hbar^3A}
\sum_{{\mathbf q}}\int\frac{d\omega}{2\pi}\left|\frac{V_{21}(q)}
{{\mathcal E}(q,\omega)}\right|^2
\left(-\frac{\partial n_B(\omega)}{\partial \omega}\right)\nonumber\\
&&\times
\Delta_2^{\alpha}({\mathbf q},{\mathbf q},\omega+i\eta,\omega-i\eta)
\Delta_1^{\gamma}(-{\mathbf q},-{\mathbf q},-\omega-i\eta,-\omega+i\eta),
\end{eqnarray}
where $\eta$ is a positive infinitesimal, $A$ is a normalization area, 
$V_{21}(q)$ is the interlayer
Coulomb interaction, ${\mathcal E}(q,\omega)$ is the screening 
function, and $n_B$ is the Bose--Einstein distribution function. 
Notice that the transconductivity
tensor is a dyadic product of the two three-body correlation functions 
${\mathbf \Delta}_2$ and ${\mathbf \Delta}_1$, which are defined by
\begin{equation}\label{triang}
\Delta_i^{\alpha}({\mathbf x},\tau;{\mathbf x'},\tau';
{\mathbf x''},\tau'')
=-\langle T_{\tau}\left\{ j_i^{\alpha}({\mathbf x},\tau)
\rho_i({\mathbf x'},\tau')\rho_i({\mathbf x''},\tau'')
\right\}\rangle,
\end{equation}
and will be referred to as the {\it triangle functions}.
The following convention for the Fourier transform has been adopted
\begin{eqnarray}
\Delta_i^{\alpha}({\mathbf x},\tau;{\mathbf x'},\tau';
{\mathbf x''},\tau'')=
\frac{1}{A^2}\sum_{{\mathbf q_1},{\mathbf q_2}}\frac{1}{(\hbar\beta)^2}
\sum_{i\omega_1,i\omega_2}
&&
e^{i{\mathbf q_1}\cdot({\mathbf x}-{\mathbf x''})+
i{\mathbf q_2}\cdot({\mathbf x'}-{\mathbf x''})}
e^{-i\omega_1(\tau-\tau'')-i\omega_2(\tau'-\tau'')}\nonumber\\
&&\times
\Delta_i^{\alpha}({\mathbf q_1}+{\mathbf q_2},{\mathbf q_2},
i\omega_1+i\omega_2,i\omega_2).
\label{ft_delta}
\end{eqnarray}
This Fourier transform convention relies on the translational invariance 
of the triangle function which applies when we consider infinite 
systems; extra caution should be exercised when considering
systems of finite extent.

\section{The triangle function}\label{triangle}
After the expansion in the interlayer interaction the 
correlation functions ${\mathbf \Delta}_1$ and 
${\mathbf \Delta}_2$ only depend on the individual layers 
and contain all microscopic details of these.
To proceed we must choose a model for the 
individual layers. 

The non-interacting electron model is 
a good approximation for the experimental systems studied 
so far\cite{gramila,sivan,hill,rubel} as long as the magnetic 
field is not too strong and the temperature not too low 
(see Introduction). Hence, we shall 
treat the individual layers as non-interacting electrons 
scattering against random impurities. 
Within this model it has been shown 
that for short range impurity potentials,\cite{kamenev,bakk}
the triangle function is proportional to 
the imaginary part of the polarization function 
\begin{equation}\label{zerores}
{\mathbf \Delta}(\pm {\mathbf q},\pm {\mathbf q},\pm 
(\omega+i\eta),\pm (\omega-i\eta))=
\frac{2\tau\hbar^2}{m^*}\ {\mathbf q}\ {\mathrm Im}\chi
(q,\omega)
\end{equation}
in absence of a magnetic field.
[The same form is recovered if electron-electron 
scattering keeps the distribution function as
a shifted Fermi-Dirac.\cite{plasmonlong}] 
Here $m^*$ is the effective 
electron mass, and the transport scattering time $\tau$ is 
assumed to be energy independent.
In the following two sections it will be shown that we can 
obtain a similar relation in the limit of $\omega_c\tau\gg 1$ 
for short ranged scattering potentials. 

Up to this point an ensemble average over impurity configurations 
has been implicitly understood. When the triangle function 
is expressed in terms of Green functions, the impurity averaging 
is accounted for by dressing the Green functions by self-energies
and including vertex functions where they connect. 
A careful account 
of impurities is necessary in the presence of a magnetic 
field because of the high Landau level degeneracy; leaving out 
impurities would lead to unphysical divergences.

The free Hamiltonian is
\begin{equation}
H=\frac{1}{2m^*}\left({\mathbf p}+e{\mathbf A}\right)^2
\end{equation}
with ${\mathbf B}={\mathbf \nabla}\times{\mathbf A}$. We choose
to work in the Landau gauge, ${\mathbf A}=(0,Bx,0)$, so that 
each eigenstate is characterized by two 
quantum numbers: $|N\rangle=|n,k\rangle$. For infinite systems, the 
eigenenergies only depend on the Landau level index $n$: 
$\varepsilon_n=(n+1/2)\hbar\omega_c$. 

\begin{figure}
\epsfxsize=7.5cm
\epsfbox{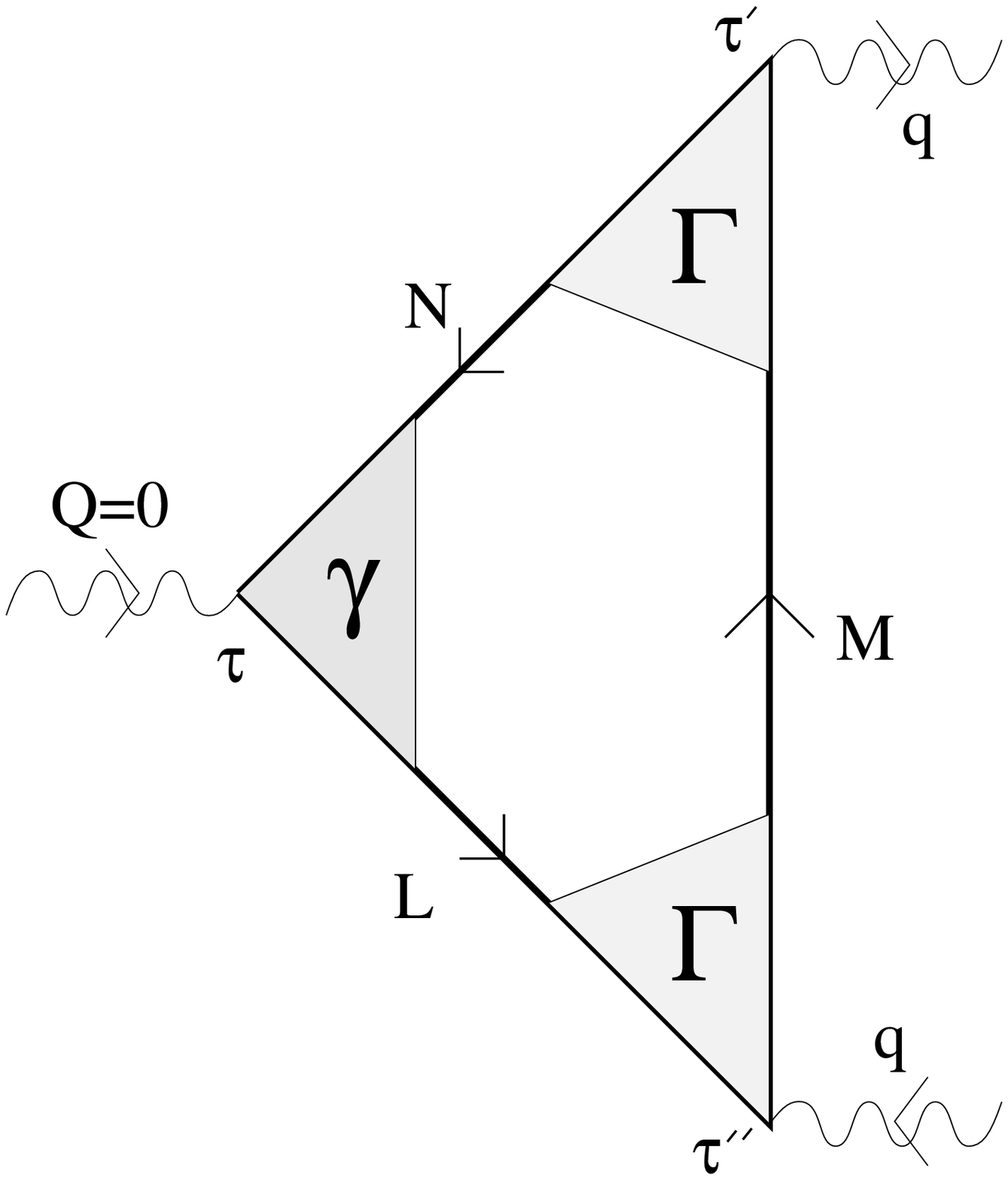}
\vspace{0.5cm}
\caption{
One of the two diagrams contributing to the triangle function 
given by Eq.~(\ref{triang}). The Green functions (solid lines) are
indexed by quantum numbers characterizing the Landau levels in the 
Landau gauge. The propagators are `dressed' by interactions with 
impurities (as shown in Fig.~\ref{self}(a)); consistent with the
Ward identity charge vertex functions $\Gamma$ and current vertex 
functions $\gamma$ are included.  Note this diagram excludes some
negligible contributions (see Fig. 4(a), 
Ref.~\protect\onlinecite{bakk}).  
We assume short-range scatterers, which implies that $\gamma$ can 
be neglected and hence $L = N\pm 1$.
}
\label{fourtri}
\end{figure}

\begin{figure}
\epsfxsize=15.5cm
\epsfbox{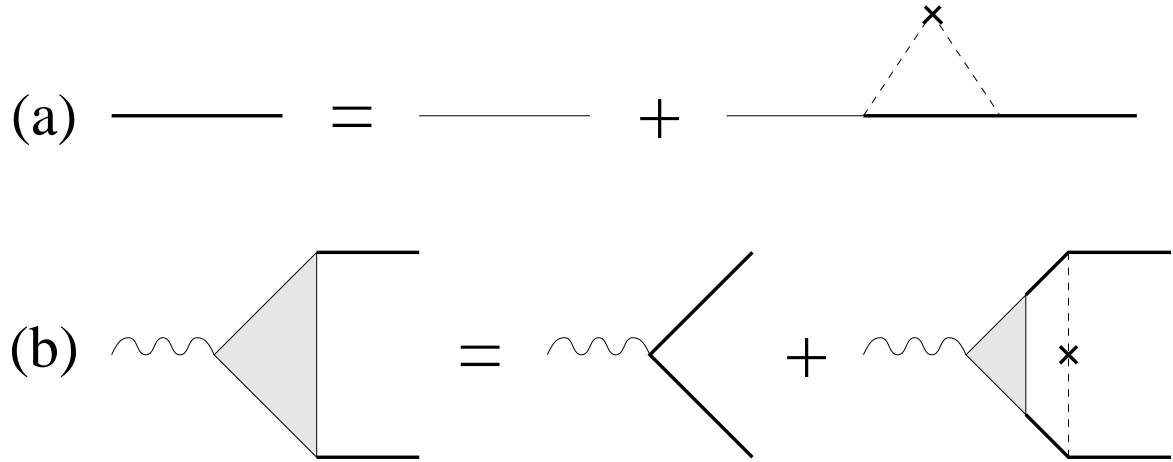}
\vspace{0.5cm}
\caption{
The impurity scattering is accounted for by the self-consistent 
Born approximation illustrated in (a). Thin lines are bare Green 
functions and thick lines are dressed Green functions. The dashed 
lines symbolizes interactions with impurities (crosses). 
Consistent with the self-consistent Born approximation, the charge 
vertex function should be taken as a ladder sum shown in (b).
}
\label{self}
\end{figure}

In terms of creation and annihilation operators, the density 
operator is 
\begin{equation}\label{density}
\rho({\mathbf x'},\tau)=\frac{1}{A}\sum_{{\mathbf q},N,M}
e^{i{\mathbf q}\cdot {\mathbf x'}}
\langle N|e^{i{\mathbf q}\cdot {\mathbf r}}|M\rangle 
c_N^\dag(\tau)c_M(\tau)
\end{equation}
and the operator for the current is given by
\begin{equation}\label{current}
{\mathbf j}(\tau)=\frac{\ell\omega_c}{A}\sum_N 
\left(
{\mathbf w}_1
\sqrt{n+1}c_{N+1}^\dag(\tau)c_N(\tau)+
{\mathbf w}_2
\sqrt{n}c_{N-1}^\dag(\tau)c_N(\tau)\right),
\end{equation}
where we have defined the two vectors, 
${\mathbf w}_1=i\hat{{\mathbf x}}+\hat{{\mathbf y}}$ and 
${\mathbf w}_2={\mathbf w}_1^*=-i\hat{{\mathbf x}}+\hat{{\mathbf y}}$
with $\hat{{\mathbf x}}$ and $\hat{{\mathbf y}}$ being unit 
vectors defining the planes of the electron layers. 
When the density and current operators are inserted in the 
expression for the triangle function, we get terms
involving the statistical average of products of 3 creation and 
3 annihilation operators. 
The Hamiltonian for impurity scattering is quadratic in creation
and annihilation operators, which means that we can use Wick's 
theorem to write the product of creation and annihilation operators
in terms of products of 3 Green functions. Inserting (\ref{density}) 
and (\ref{current}) in Eq.~(\ref{triang}) we get two connected 
diagrams.  One of these is given by (see Fig.\ \ref{fourtri})
\begin{eqnarray}\label{oneterm}
\frac{-1}{A^3}\sum_{N,N',M,{\bf q}} && e^{i{\bf q}\cdot ({\bf x'}-{\bf x''})} 
\Gamma({\bf q},N,M,\tau-\tau',\tau'-\tau'')G(N,\tau-\tau')\nonumber\\
&&
\gamma^{\alpha}({\bf Q}=0,N',N,\tau''-\tau,\tau-\tau')G(N',\tau''-\tau)\nonumber\\
&&
\Gamma(-{\bf q},M,N',\tau'-\tau'',\tau''-\tau)G(M,\tau'-\tau'')
\end{eqnarray}
where a current vertex $\gamma$ and charge vertices $\Gamma$ have been
included to take account of impurity scattering; in the other diagram
the direction of the arrows is reversed.
The scattering of electrons against impurities is evaluated in the 
self-consistent Born approximation. The self-energy diagram is shown in 
Fig.\ \ref{self}a.
In the limit of $\omega_c\tau\gg 1$ Ando and Uemura\cite{ando} have 
shown that the self-energy is given by
\begin{equation}\label{self-energy}
\hbar\Sigma^{r,a}(n,\epsilon)=\frac{\hbar\epsilon-\varepsilon_n}{2}
-\frac{\Gamma_0}{2}\sqrt{\left(\frac{\hbar\epsilon-\varepsilon_n}
{\Gamma_0}\right)^2-1}
\end{equation}
where $\Gamma_0^2=(2/\pi)\hbar\omega_c(\hbar/\tau)$, 
$\tau$ being the transport scattering time at zero magnetic field. 
The imaginary part of Eq.~(\ref{self-energy}) is taken as negative
(positive) for the retarded (advanced) function. 
The width of the Landau level, $2\Gamma_0$, is independent of the 
Landau level index if the range of the scattering potential is 
smaller than the magnetic length. 
The choice of self-energy diagram implies, by a Ward identity, 
a specific choice of vertex functions.  
Born approximation for the self-energy implies that we 
must sum ladder diagrams for the vertex functions. In the limit 
of short range scattering potential the contribution from the ladder 
sum to the current vertex function can be neglected,\cite{ando}
{\it i.e.} $\gamma$ can be approximated by a bare current vertex and
hence $N' = N\pm 1$ in Eq.\ (\ref{oneterm}).
For the charge vertex, on the other hand, the ladder sum is 
important.  Fig.\ \ref{self}b shows the diagrams corresponding to the 
following integral equation for $\Gamma$:
\begin{eqnarray}\label{vertex}
\Gamma({\mathbf q},n,m,i\epsilon_1,i\epsilon_2)
&=&
f_{n,m}({\mathbf q})+\frac{\rho_{{\mathrm imp}}}{\hbar^2}
\sum_{a,b}\int\!\frac{d{\mathbf k}}{(2\pi)^2}U^2(k)
f_{b,m}({\mathbf k})f_{n,a}(-{\mathbf k})\nonumber\\
&&
\times {\mathcal G}(a,i\epsilon_1){\mathcal G}(b,i\epsilon_2)
e^{i\ell^2(k_xq_y-k_yq_x)}
\Gamma({\mathbf q},a,b,i\epsilon_1,i\epsilon_2),
\end{eqnarray}
where $\rho_{{\mathrm imp}}$ is the density of impurities, 
$U(k)$ is the impurity potential, and ${\mathcal G}$ is a 
Matsubara Green function. The bare charge vertex is given by 
\begin{equation}
f_{n,m}({\mathbf q})=\left\{
\begin{array}{ll}
e^{-(\ell q/2)^2}\sqrt{\frac{m!}{n!}2^{m-n}}\left(-\ell(iq_x+q_y)
\right)^{n-m}L_m^{n-m}((\ell q)^2/2) & \ \ ,m\leq n \\
e^{-(\ell q/2)^2}\sqrt{\frac{n!}{m!}2^{n-m}}\left(-\ell(iq_x-q_y)
\right)^{m-n}L_n^{m-n}((\ell q)^2/2) & \ \ ,n\leq m 
\end{array}
\right.
\end{equation}
where $L_n^m$ are the Laguerre polynomials and 
$\ell=\sqrt{\hbar/eB}$ is the magnetic length. 

In terms of dressed Matsubara Green functions and vertex functions, 
the expression for the triangle function is
\begin{equation}
{\mathbf \Delta}({\mathbf q},{\mathbf q},i\Omega+i\omega,i\omega)
=\frac{-\omega_c}{2\pi\ell\hbar\beta}\sum_{n,m,i\omega_1}
\sqrt{n}\left({\mathbf w}_1{\mathcal F}_1 + {\mathbf w}_2{\mathcal F}_2
\right)\;,
\end{equation}
where
\begin{eqnarray}
{\mathcal F}_1&=&\Gamma({\mathbf q},n-1,m,i\omega_1,i\omega_1+i\omega)
\Gamma(-{\mathbf q},m,n,i\omega_1+i\omega,i\omega_1-i\Omega)\nonumber\\
&\times&
{\mathcal G}(n,i\omega_1-i\Omega){\mathcal G}(m,i\omega_1+i\omega)
{\mathcal G}(n-1,i\omega_1)\nonumber\\
&+&
\Gamma({\mathbf q},m,n,i\omega_1-i\omega,i\omega_1)
\Gamma(-{\mathbf q},n-1,m,i\omega_1+i\Omega,i\omega_1-i\omega)\nonumber\\
&\times&
{\mathcal G}(n,i\omega_1){\mathcal G}(m,i\omega_1-i\omega)
{\mathcal G}(n-1,i\omega_1+i\Omega)
\end{eqnarray}
and
\begin{eqnarray}
{\mathcal F}_2&=&\Gamma({\mathbf q},n,m,i\omega_1,i\omega_1+i\omega)
\Gamma(-{\mathbf q},m,n-1,i\omega_1+i\omega,i\omega_1-i\Omega)\nonumber\\
&\times&
{\mathcal G}(n,i\omega_1){\mathcal G}(m,i\omega_1+i\omega)
{\mathcal G}(n-1,i\omega_1-i\Omega)\nonumber\\
&+&
\Gamma({\mathbf q},m,n-1,i\omega_1-i\omega,i\omega_1)
\Gamma(-{\mathbf q},n,m,i\omega_1+i\Omega,i\omega_1-i\omega)\nonumber\\
&\times&
{\mathcal G}(n,i\omega_1+i\Omega){\mathcal G}(m,i\omega_1-i\omega)
{\mathcal G}(n-1,i\omega_1)\;.
\end{eqnarray}
The summation over Matsubara frequencies $i\omega_1$ can be carried
out as a contour integration. The function ${\mathbf \Delta}
({\mathbf q},{\mathbf q},\omega+i\eta,\omega-i\eta)$ is then
obtained by letting $i\Omega+i\omega\rightarrow\Omega+\omega+i\eta$,
$i\Omega\rightarrow\Omega+i\eta$, 
and $i\omega\rightarrow\omega-i\eta$.\cite{bakk}
In the static limit, $\Omega\rightarrow 0$, the result is
\begin{eqnarray}\label{fourp}
{\mathbf \Delta}({\mathbf q},{\mathbf q},\omega+i\eta,\omega-i\eta)
&=&
\frac{\hbar\omega_c}{2\pi\ell}\sum_{n,m}\sqrt{n}\int\frac{d\epsilon}
{2\pi i}n_F(\epsilon)\nonumber\\
&&\times
\Big[{\mathbf w}_1\Big(P({\mathbf q},n,m,\epsilon,\epsilon+\omega)
+P(-{\mathbf q},n,m,\epsilon,\epsilon-\omega)\Big)\nonumber\\
&&
\ \ \ -{\mathbf w}_2\Big(
P^*({\mathbf q},n,m,\epsilon,\epsilon+\omega)
+P^*(-{\mathbf q},n,m,\epsilon,\epsilon-\omega)\Big)
\Big]
\end{eqnarray}
with
\begin{eqnarray}\label{pexpr}
P&&({\mathbf q},n,m,\epsilon,\epsilon+\omega)=\nonumber\\
&&
\Gamma^{+-}({\mathbf q},n,m,\epsilon,\epsilon+\omega)
\Gamma^{--}(-{\mathbf q},m,n-1,\epsilon+\omega,\epsilon)
G^r(n,\epsilon)G^{a}(m,\epsilon+\omega)G^{a}(n-1,\epsilon)\nonumber\\
&&
-\Gamma^{--}({\mathbf q},n,m,\epsilon,\epsilon+\omega)
\Gamma^{--}(-{\mathbf q},m,n-1,\epsilon+\omega,\epsilon)
G^{a}(n,\epsilon)G^{a}(m,\epsilon+\omega)G^{a}(n-1,\epsilon)\nonumber\\
&&
+\Gamma^{++}({\mathbf q},n,m,\epsilon-\omega,\epsilon)
\Gamma^{+-}(-{\mathbf q},m,n-1,\epsilon,\epsilon-\omega)
G^r(n,\epsilon-\omega)G^{r}(m,\epsilon)G^{a}(n-1,\epsilon-\omega)\nonumber\\
&&
-\Gamma^{+-}({\mathbf q},n,m,\epsilon-\omega,\epsilon)
\Gamma^{--}(-{\mathbf q},m,n-1,\epsilon,\epsilon-\omega)
G^r(n,\epsilon-\omega)G^{a}(m,\epsilon)G^{a}(n-1,\epsilon-\omega)\nonumber\\
&&
+\Gamma^{++}({\mathbf q},n,m,\epsilon,\epsilon+\omega)
\Gamma^{++}(-{\mathbf q},m,n-1,\epsilon+\omega,\epsilon)
G^r(n,\epsilon)G^{r}(m,\epsilon+\omega)G^{r}(n-1,\epsilon)\nonumber\\
&&
-\Gamma^{++}({\mathbf q},n,m,\epsilon,\epsilon+\omega)
\Gamma^{+-}(-{\mathbf q},m,n-1,\epsilon+\omega,\epsilon)
G^r(n,\epsilon)G^{r}(m,\epsilon+\omega)G^{a}(n-1,\epsilon).
\end{eqnarray}
The plus and minus signs attached to the vertex functions indicate 
the signs of the imaginary infinitesimals that should be added to 
the frequency arguments. 
{}From (\ref{fourp}) it can be realized that ${\mathbf \Delta}$ is
a vector with purely real components. 
Other general properties of ${\bf\Delta}$ are given in Appendix
\ref{app:deltaprop}.

\section{Triangles to bubbles}\label{tribub}

We now show that in the limit of short range
scattering potentials and for $\omega_c\tau\gg 1$, we can 
express the triangle function in terms ${\mathbf q}$ and 
the imaginary part of the proper polarization function.

\begin{figure}
\epsfxsize=13.5cm
\epsfbox{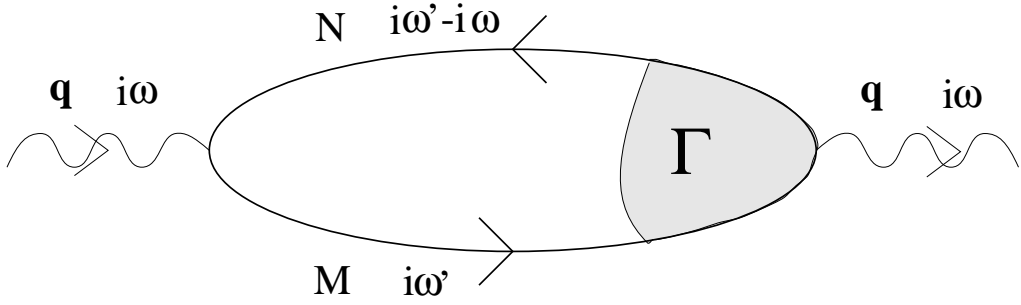}
\vspace{0.5cm}
\caption{
The density--density correlation function. Solid lines are dressed 
Green functions and the shaded area is the charge vertex function.
The proper polarization function is obtained by doing analytical 
continuation, $i\omega\rightarrow\omega+i\eta$. The explicit expression
is given in Eq.~(\ref{sus}).
}
\label{polar}
\end{figure}

The proper polarization function is obtained by analytical 
continuation of the density--density correlation function 
shown in Fig.~\ref{polar}.
{}From the structure it is seen, that it involves two Green 
functions $G$, one bare vertex $f$, and one vertex function 
$\Gamma$; symbolically $\chi\sim f\,G\,G\,\Gamma$. The triangle 
function, on the other hand, involves products of three 
Green functions and two vertex functions; symbolically 
$\Delta\sim G\,G\,G\,\Gamma\,\Gamma$. To reduce the triangle function
to the polarization function we must therefore reduce 
three Green functions to two, and two vertex functions to
one vertex function and a bare vertex. Furthermore we 
must introduce a factor of ${\mathbf q}$ if we want an
expression similar to Eq.~(\ref{zerores}). Symbolically the task 
is to do the simplification: 
$G\,G\,G\,\Gamma\,\Gamma \longrightarrow {\mathbf q}\,G\,G\,\Gamma\,f$,
which we shall now proceed to carry out.

The key to the problem is to notice that in the expression
(\ref{pexpr}) two of the Green functions and both the vertex 
functions in the product $G\,G\,G\,\Gamma\,\Gamma$ have neighboring 
Landau level indices, $n$ and $n-1$.
The retarded and advanced Green functions are given by
$G^{r,a}(n,\epsilon)=[\epsilon-\frac{1}{\hbar}\varepsilon_n-
\Sigma^{r,a}(n,\epsilon)]^{-1}$.
Using the identity $1/AB=(1/A-1/B)/(B-A)$
we get
\begin{equation}\label{appp}
G^r(n,\epsilon)G^{a}(n-1,\epsilon)=\frac{G^r(n,\epsilon)-G^{a}(n-1,\epsilon)}
{\omega_c+\Sigma^r(n,\epsilon)-\Sigma^{a}(n-1,\epsilon)}.
\end{equation}
In the limit $\omega_c\tau\gg 1$ the self-energies 
can be neglected compared to the cyclotron frequency, and we can approximate
\begin{equation}\label{aprd}
G^r(n,\epsilon)G^{a}(n-1,\epsilon)
\simeq\frac{G^r(n,\epsilon)-G^{a}(n-1,\epsilon)}{\omega_c}.
\end{equation}
We have thus reduced products of three Green functions to products of 
two Green functions.

To reduce the product of two vertex functions to one vertex function and 
one bare vertex is more involved and therefore deferred to Appendix 
\ref{appvertex} where it is shown that we can do the approximation
in leading order of $(\omega_c\tau)^{-1}$
\begin{eqnarray}\label{approx}
\Gamma^{+-}&&({\mathbf q},n,m,\epsilon,\epsilon+\omega)
\Gamma^{--}(-{\mathbf q},m,n-1,\epsilon+\omega,\epsilon)
G^{r}(n,\epsilon)G^{a}(m,\epsilon+\omega)G^{a}(n-1,\epsilon)\nonumber\\
&&
\simeq\frac{1}{\omega_c}\Gamma^{+-}({\mathbf q},n,m,\epsilon,\epsilon+\omega)
f_{m,n-1}(-{\mathbf q})
G^r(n,\epsilon)G^{a}(m,\epsilon+\omega)\nonumber\\
&&\ \ 
-\frac{1}{\omega_c}
f_{n,m}({\mathbf q})\Gamma^{--}(-{\mathbf q},m,n-1,\epsilon+\omega,\epsilon)
G^{a}(m,\epsilon+\omega)G^{a}(n-1,\epsilon).
\end{eqnarray}
Notice that the full vertex function with same Landau indices as the Green functions 
is retained, and that the signs of the infinitesimals in the vertex function
naturally follow the signs of the infinitesimal on the Green functions that 
it is multiplying. 
In the above expression there seems to be a mismatch between the Landau level indices
of the Green functions and the Landau level indices of the bare vertices $f_{n,m}$
multiplying them. Matching indices are recovered by the identity 
\begin{equation}\label{ident}
\sqrt{n+1}f_{n+1,m}({\mathbf q})-\sqrt{m}f_{n,m-1}({\mathbf q})
=\frac{-\ell(iq_x+q_y)}{\sqrt{2}}f_{n,m}({\mathbf q}),
\end{equation}
which also introduces ${\mathbf q}$ into the expression. 
The square root factors in Eq.~(\ref{ident}) come from the bare 
current vertices (see Eq.~(\ref{current})). With these 
approximations it is a matter of simple manipulations to reach 
the following relation
\begin{equation}\label{mainres}
{\mathbf \Delta}(\pm {\mathbf q},\pm {\mathbf q},\pm (\omega+i\eta),
\pm(\omega-i\eta))=
\mp\frac{2\hbar^2}{eB^2}\,{\mathbf q}\times{\mathbf B}\ 
{\mathrm Im}\chi(q,\omega),
\end{equation}
which is valid for $\omega_c\tau\gg 1$.
Note that for $B=0$, 
${\mathbf \Delta}(\pm {\mathbf q},\pm {\mathbf q},\pm (\omega+i\eta)$
have the same sign.
Here the (irreducible) polarization function is
\begin{eqnarray}\label{sus}
\chi&&(q,\omega)=\frac{1}{\hbar\pi\ell^2}
\sum_{n,m}\int
\frac{d\epsilon}{2\pi i}n_F(\epsilon)\nonumber\\
&&
\times\Big[
G^r(m,\epsilon+\omega)f_{n,m}({\mathbf q})\nonumber\\
&&\ \ \ \ \ \ \ 
\times\left\{G^{a}(n,\epsilon)\Gamma^{+-}(-{\mathbf q},m,n,\epsilon+\omega,\epsilon)
-G^r(n,\epsilon)
\Gamma^{++}(-{\mathbf q},m,n,\epsilon+\omega,\epsilon)\right\}
\nonumber\\
&&\ \ \ 
+G^{a}(m,\epsilon-\omega)f_{m,n}({\mathbf q})\nonumber\\
&&\ \ \ \ \ \ \ 
\times\left\{G^{a}(n,\epsilon)\Gamma^{--}(-{\mathbf q},n,m,\epsilon,\epsilon-\omega)
-G^r(n,\epsilon)\Gamma^{+-}(-{\mathbf q},n,m,\epsilon,\epsilon-\omega)
\right\}\Big].
\end{eqnarray}

To obtain the transresistivity we must know the single--layer 
resistivities $\tensor{\rho_{ii}}$ [see Eq.~(\ref{basic})].
For isotropic systems they have the generic structure 
\begin{equation}\label{single}
\tensor{\rho_{ii}}=\rho_{0i}
\left(
\begin{array}{cc}
a_i  &  b_i \\
-b_i  &  a_i
\end{array}
\right)
\end{equation}
with $\rho_{0i}=m_i^*/(n_ie^2\tau_i)$ being the resistivity in zero 
magnetic field. Combining Eqs.\ (\ref{basic}), (\ref{mainres}) and 
(\ref{single}) we find that for $b\gg a$ the transresistivity 
tensor is diagonal in cartesian coordinates; the diagonal elements given by
\begin{equation}\label{rho}
\rho_{21}^{xx}=\left[\frac{b_1b_2}{(\omega_c\tau)_1(\omega_c\tau)_2}\right]
\frac{(-\hbar^2)}{4e^2n_1n_2k_BT}\frac{1}{A}\sum_{{\mathbf q}} q^2
\int\frac{d\omega}{2\pi}
\left| \frac{V_{21}(q)}{{\mathcal E}(q,\omega)} \right|^2
\frac{\mathrm{Im}\chi_1({\mathbf q},\omega)\mathrm{Im}\chi_2({\mathbf q},\omega)}
{\sinh^2(\hbar\omega/2k_BT)},
\end{equation}
A semiclassical treatment of $\tensor{\rho}_{ii}$ yields 
$b_i=(\omega_c\tau)_i$, so that the term in the square bracket 
above is unity. We assume this to be the case in our numerical 
evaluations. As the quantum Hall regime is approached, 
$b_i/(\omega_c\tau)_i$ starts deviating from 1; however this deviation 
does not change the main features of the numerical results presented 
in Sec.\ \ref{num}.

\subsection{A conjecture: generalization to arbitrary $B$-field}\label{res}
In the previous subsection it was shown, that when the individual
layers are treated as non-interacting electrons scattering 
against short-range impurities, the triangle function is 
related to the imaginary part of the polarization function 
in the limit of $\omega_c\tau\gg 1$.  Work has also been done
in the small magnetic field limit $\omega_c\tau\ll 1$.\cite{kamenev,halldrag}
We now discuss a conjecture for the generalization of the 
expression for ${\bf\Delta}$ for arbitrary magnetic field strengths
which extrapolates between the weak and strong field limits.

For zero magnetic field  the triangle 
function is proportional to the imaginary part of the 
polarization function when the impurity scattering time is 
independent of energy (see Eq.~(\ref{zerores})).\cite{bakk}
One case of an energy independent transport time is when the 
range of the impurity potential is short compared to the 
Fermi wavelength. Likewise, for high magnetic fields we found 
that a prerequisite for a simple relation between 
${\mathbf \Delta}$ and Im$\chi$ is short ranged scatterers. 
The task in this section is to 
bridge the gap between zero and high magnetic field.
In order to do this we first observe that 
only two vectors can be constructed in the $xy$-plane, namely 
${\mathbf q}$ and ${\mathbf q}\times{\mathbf B}$. The triangle 
function is therefore of the form
\begin{equation}
{\mathbf \Delta}(\pm{\mathbf q},\pm{\mathbf q},\pm (\omega+i\eta),
\pm (\omega-i\eta))=\Delta_{\|}(q,\omega,B){\mathbf\hat{q}}\pm 
\Delta_{\perp}(q,\omega,B) {\mathbf\hat{q}}\times {\mathbf \hat{B}}\;,
\label{delta_form}
\end{equation}
where the carets denote unit vectors. 
Knowledge of the zero and high magnetic field limits, results 
from a semiclassical analysis,\cite{halldrag}
and a perturbational calculation\cite{kamenev} suggest the following 
conjecture for the form of the triangle function, valid for short 
range scatterers but for arbitrary magnetic field strength: 
\begin{equation}\label{conjecture}
{\mathbf \Delta}_i(\pm{\mathbf q},\pm{\mathbf q},\omega\pm i\eta,
\omega\mp i\eta)=\frac{2\tau_i\hbar^2}{m^*_i}\left(
\frac{1}{1+\alpha_i^2}{\mathbf q} \mp \frac{\alpha_i}{1+\alpha_i^2}
{\mathbf q}\times {\mathbf \hat{B}}\right)
{\mathrm Im}\chi_i(q,\omega,B),
\end{equation}
where $\alpha_i(B)$ is a parameter to be determined.
The magnetic field has been added as an argument of the polarization 
function to emphasize that it should be evaluated in the presence 
of the magnetic field. 

The $\alpha_i(B)$ should be chosen so that Eq.\ (\ref{conjecture})  
is consistent with known results.  One such empirical result is that
so far no experiment has ever observed Hall drag.\cite{privcom} 
If one {\sl assumes} that Hall drag is absent 
({\it i.e.}, $\rho_{21}^{xy} = 0$), then 
\begin{equation}
\alpha_i = \frac{b_i}{a_i}.
\label{choice}
\end{equation}
With the above choice one obtains in the low-field Drude limit 
$\alpha_i = (\omega_c\tau)_i$, which is consistent with semiclassical 
low-field results.\cite{halldrag}
Furthermore, we note it reproduces the result obtained using the 
memory-functional formalism of Ref.~\onlinecite{zheng}.

We can illustrate the plausibility of a vanishing Hall-drag by
a very simple argument.
The electrons in layer 2 are influenced by two forces which must add 
up to zero because there is no current in layer 2, {\it i.e.}
\begin{equation}
0={\mathbf F}_{12}+(-e)\left( {\mathbf E}_2+\langle{\mathbf v}_2
\rangle\times{\mathbf B}\right)\;,
\end{equation}
where ${\mathbf F}_{12}$ is the force from the electrons in layer 1,
${\mathbf E}_2$ is the induced electric field, and 
$\langle{\mathbf v}_2\rangle$ is 
the average velocity of the electrons in layer 2. When no current is 
allowed to flow in layer 2, $\langle{\bf v}_2\rangle=0$. Furthermore, since 
the Coulomb force is radial, ${\mathbf F}_{12}$ 
must be parallel to ${\mathbf J}_1$ and hence the measured electric 
field is parallel to the driving current, {\it i.e.} no Hall drag. 

However, the above argument is not as general as it 
seems,\cite{halldrag} and to second-order in the screened 
interlayer interaction, Hall-drag can occur in cases where 
the band-structure is anisotropic, which breaks in-plane inversion
symmetry and when the intralayer scattering time is energy-dependent,
which does not allow a simple description based on
the polarization function alone.\cite{plasmonlong,halldrag}
Furthermore, higher-order correlation effects
such as those found in an electron-hole system with bound 
pairs\cite{vignale}
may lead to a finite Hall-drag even in the absence of the 
conditions mentioned above.

\section{Numerical results}\label{num}
The formula for the transresistivity Eq.\ (\ref{rho}) must 
be evaluated numerically. We focus on the dependence on
magnetic field strength and temperature. 
As a model for the dielectric function, we adopt the random
phase approximations in which
\begin{equation}
{\mathcal E}(q,\omega)=[1-\chi(q,\omega)
V_{11}(q)]^2-[\chi(q,\omega)V_{21}(q)]^2,
\end{equation}
where $V_{11}(q)$ is the intralayer- and 
$V_{21}(q)$ is the interlayer Coulomb interaction (we 
have assumed that $\chi_1=\chi_2\equiv\chi$, {\it i.e.} 
identical layers).

The polarization function enters both directly in the 
Eq.\ (\ref{rho}) and indirectly through the dielectric function.
The general expression for $\chi(q,\omega)$ is given 
in Eq.\ (\ref{sus}). To make the numerical evaluations 
tractable, it is necessary to make an approximation
for the vertex functions, which in general are given 
by the integral equation (\ref{vertex}). 
In Appendix \ref{appvertex} we show that when the 
Landau levels are clearly resolved, we can approximate
\begin{equation}
\Gamma({\mathbf q},n,m,i\epsilon+i\omega,i\epsilon)=
\frac{f_{n,m}({\mathbf q})}{1-(\Gamma_0/2\hbar)^2
{\mathcal I}(q,n,m)
{\mathcal G}(n,i\epsilon+i\omega){\mathcal G}(m,i\epsilon)}
\end{equation}
with ${\mathcal I}(q,n,m)$ given by Eq.~(\ref{iii}). 
This approximation is consistent with the assumptions
under which we derived Eq.\ (\ref{rho}) and makes the numerical 
evaluation tractable. 

In order to avoid unphysical jumps in
the chemical potential we must improve the self-consistent Born
approximation 
(which leads to a vanishing density of states outside the
Landau bands), and
we use a Gaussian the density of states
derived by Gerhardts,\cite{gerhardts}
\begin{equation}
g_{m,\sigma}(\varepsilon)=\frac{\sqrt{2/\pi}}{2\pi\ell^2\Gamma_0}
\exp\left[-2\left(\frac{\varepsilon-\varepsilon_m}
{\Gamma_0}\right)^2\right]
\end{equation}
where $\sigma$ denotes spin. The chemical
potential $\mu(B,T)$ is determined implicitly
by requiring the density $n$ to be given by
\begin{equation}
n=\sum_{m,\sigma}\int_0^{\infty}\! d\varepsilon\ n_F(\varepsilon-\mu)
g_{m,\sigma}(\varepsilon).
\end{equation}
This model has a finite range of magnetic fields where the 
density of extended states at the Fermi energy is suppressed, 
simulating the effect of localized states between the Landau bands,
which  are needed to obtain
quantum Hall plateaus with a finite width. 
However, the quantitative details of localization, such as the
critical properties of the metal-insulator transition, are
not included in this simple model.

With these approximations we evaluate the transresistivity given
by Eq.~(\ref{rho}) as a function of magnetic field and temperature. 
For simplicity we consider two identical electron layers of densities
$n_1=n_2=3\times10^{15}$ m$^{-2}$ corresponding to a Fermi temperature
of $T_F\simeq 120$ K. The center--to--center distance, $d$, is 
chosen to be 800 \AA\  and the well-widths are taken to be 200 \AA.
The Landau level width is dependent on the transport scattering time,
$\tau$, which we determine by choosing a mobility, $e\tau/m^*$, of 
25 m$^2$/Vs. 
The temperature dependence of the scattering rate -- which for 
simplicity is neglected in what follows -- will eventually lead 
to a violation of the requirement $\omega_c\tau\gg 1$, and would 
set the upper temperature limit of the validity of the numerical 
evaluations.

\subsection{Magnetic field dependence}
We focus on a magnetic field regime where the Landau levels are fully resolved.
For simplicity we neglect spin splitting. (Because of spin degeneracy, 
note that filling factor $\nu$ = odd number 
corresponds to a half-filled Landau level, whereas filled Landau levels
have $\nu$ = even number.) 
Experimentally there is 
a large regime where doubly occupied, clearly 
distinguishable Landau levels can be observed.
For half-filled Landau levels, the density 
of states $g$ is enhanced over the $B=0$ value: 
$g=g_0\sqrt{2\omega_c\tau/\pi}$ where $g_0=m^*/\pi\hbar^2$. This is 
due to the large degeneracy of the Landau levels and implies that 
there are more 
available states close to the Fermi energy for the electrons to 
scatter into. Consequently a general enhancement of the transresistivity 
should be expected in a magnetic field. Experimentally 
the transresistivity has been found to increase as the square of the 
magnetic field as long as the Landau levels are not resolved.\cite{hill} 
When the Landau levels get resolved the picture is more complicated.
At even filling factors the 
density of states is suppressed; an excitation gap develops, and the 
transresistivity should vanish as a result. These two expectations 
are both 
based on considerations of the density of states, {\it i.e.} the phase-space
available for the interlayer {\it e--e} scattering. 

The screening of the double layer system is strongly affected by 
the density of states and thus also strongly dependent on the 
magnetic field. As the density of states at the Fermi level becomes smaller 
the electron layers lose their ability to screen and hence the effective 
interlayer interaction is enhanced. 

\begin{figure}
\epsfxsize=12cm
\epsfbox{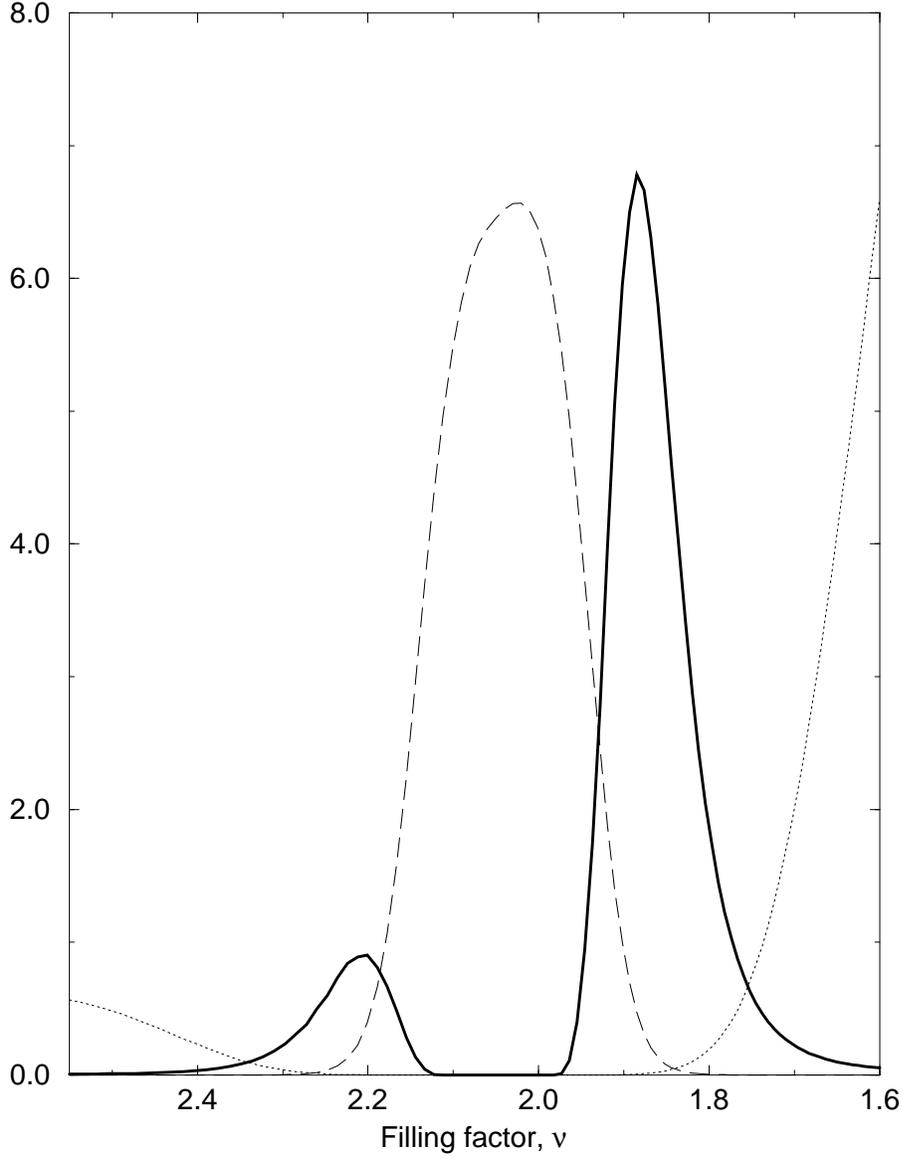}
\vspace{0.5cm}
\caption{
The two ingredients `effective interaction' and `phase-space' as 
a function of filling factor for fixed $(q,\omega)=(0.2/k_F,0.001\varepsilon_F/\hbar)$
where $\varepsilon_F$ is the Fermi energy. The temperature is given by 
$T/T_F=0.01$. 
The dashed curve is the square of the effective interaction 
$|V_{21}(q)/{\mathcal E}(q,\omega)|^2$ 
in units of $g_0^{-2}=(\pi\hbar^2/m^*)^2$, the dotted curve is 
'the phase--space' $[{\mathrm Im}\Pi(q,\omega)/\sinh(\hbar\omega/2k_BT)]^2$ 
in units of $(10g_0)^2$. The solid line is (80 times) the product of the two. 
}
\label{product}
\end{figure}

The resulting transresistivity 
can qualitatively be understood as a product of the available 
phase-space and the effective interaction. In Fig.\ \ref{product} 
we plot $|V_{21}(q)/{\mathcal E}(q,\omega)|^2$ and 
[Im$\chi(q,\omega)/\sinh(\hbar\omega/2k_BT)]^2$ as a function of 
filling factor together with the product of the two functions for
a given $q$ and $\omega$. 
The maxima of the product occur at magnetic fields for which the 
filling factor is slightly above or below an even integer 
(where an integral number of Landau levels are filled).

\begin{figure}
\epsfxsize=10cm
\epsfbox{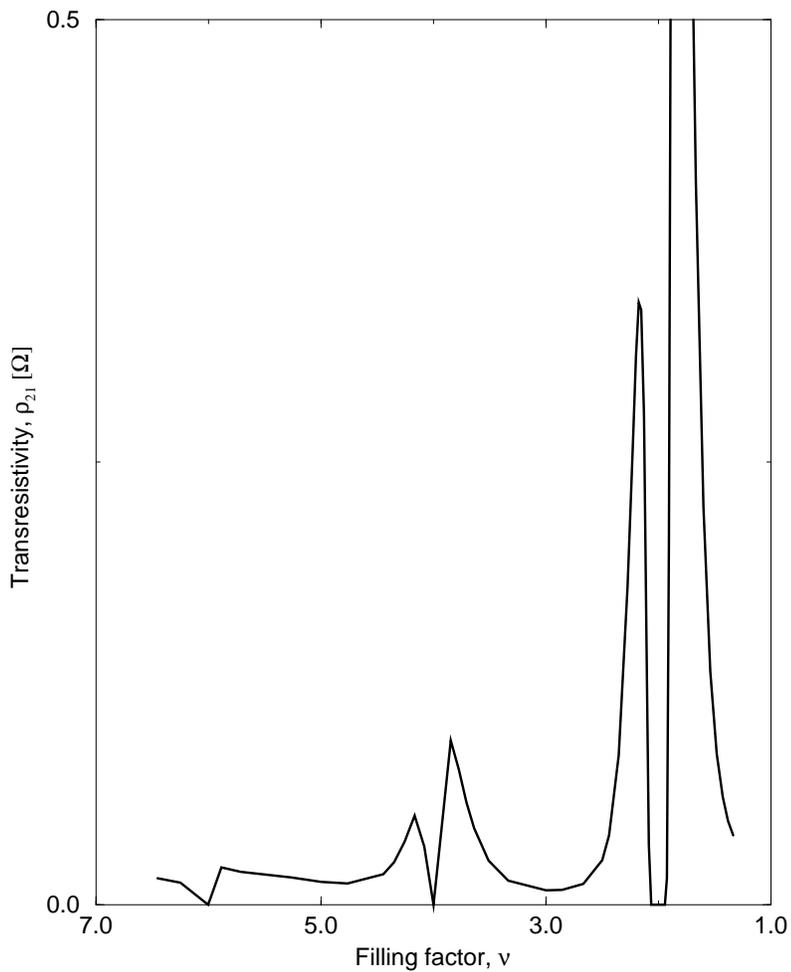}
\vspace{0.5cm}
\caption{
The transresistivity as a function of filling factor for temperature $T=1.2$ K and 
interlayer distance $d=800$ \AA. The density of the electron gases is 
$n_1=n_2=3\times 10^{15}\ {\rm m}^2$ and the mobility is 25 m$^2$/Vs. Spin 
splitting has been neglected. The transresistivity shows a twin-peak 
structure: as the filling factor is changed from an odd value (where the 
highest Landau level is half filled) towards an even value, the transresistivity
goes through a maximum before it gets suppressed. ($\rho_{21}$ has 
a maximum of 1.16 $\Omega$ at $\nu=1.8$ --- out of the range of the 
plot). 
}
\label{rhoofnu}
\end{figure}

Fig.\ \ref{rhoofnu} shows the transresistivity as a function of magnetic field. 
At odd filling factors $\rho_{21}$ is enhanced (by a factor of $\sim 100$ 
at $\nu=3$ depending on the temperature) over the zero field value as expected. 
As the magnetic field is changed from an odd filling factor towards 
an even, we find that the transresistivity {\it increases} before 
it eventually gets suppressed when the chemical potential enters the 
excitation gap. This unusual behavior is explained by the competition 
of available phase-space and effective interaction.

When comparing this theory with experiments, one faces the complication 
of spin splitting which is present in real systems.
Thus, a double peak in an experimental $\rho_{21}$ 
\cite{hill,rubel} may be due to two partially overlapping
single-peaked structures; this is the interpretation
of Ref.~\onlinecite{hill}.
However, Rubel {\it et al.}\cite{rubel} have 
shown experimentally that there is a regime of magnetic fields 
($\nu=$ 6--15) where the single-layer longitudinal resistivity shows 
no spin splitting while the transresistivity has a clear twin-peak
structure.  On the other hand, their data at higher
magnetic fields includes spin resolved structures that do {\it not}
show the predicted double-peak structure.  An improved theory,
which includes spin-splitting would clearly be desirable.

\subsection{Temperature dependence}\label{tempdep}
We will discuss two regimes of temperature which show interesting 
behavior and which yield information about the polarization 
function and the effective interaction. 

{}From general properties of density-response functions\cite{forster} 
it follows that for $q^{-1}$ 
larger than the smallest relevant length scale ( = elastic mean free path
for $B=0$, and $\ell$ for sufficiently large $B$) and $\omega$ smaller than the 
inverse scattering time, the polarization function assumes a diffusive 
form
\begin{equation}
\chi(q,\omega)=-g(\mu)\frac{Dq^2}{Dq^2-i\omega},
\label{chi_diff}
\end{equation}
where $D$ is the diffusion constant. For high magnetic fields 
the magnetic length $\ell=\sqrt{\hbar/eB}$
is smaller than typical interlayer distances $d$ 
($\ell=180$ \AA\ for $B=2$ T).
The dominant contribution to the $q$-integral in (Eq.\ \ref{rho}) 
comes from $q\leq 1/d$, and the $\omega$-integral is 
dominated by contributions form $\hbar\omega
\lesssim k_BT$. Hence, if the thermal energy is smaller than $\hbar/\tau$, 
the diffusive form of $\chi$ prevails and we should therefore expect 
$\rho_{21}\sim T^2\ln T$ as shown by Zheng and MacDonald.\cite{zheng}
Numerical evaluations showing the $T^2\ln T$ dependence of $\rho_{21}$ 
were presented in a previous publication.\cite{prl} 
The temperature below which the diffusive behavior prevails is given 
by $k_BT_{{\mathrm diff}}\approx\hbar/\tau$ and is therefore 
sample specific. For our choices of parameters, we find numerically 
that the diffusive behavior sets in at $T=0.4$ K.
The $T^2\ln T$ is a direct consequence of the diffusive form Eq.\
(\ref{chi_diff}), which emerges from Eq.~(\ref{sus}) by virtue of the
use of the self-consistent Born approximation for the vertex 
correction $\Gamma$.  As mentioned in the Introduction, other temperature 
dependences are conceivable depending on the filling factor, the 
temperature regime, and the mobility of the sample.

For temperatures higher than $T_{{\mathrm diff}}$, the dominant 
contributions to the $\omega$--integral come from $\omega>1/\tau$. 
In this regime both the real and imaginary part of the polarization 
function are strongly frequency dependent; consequently the same is 
true for the effective interaction $V_{21}(q)/{\mathcal E}
(q,\omega)$. Wu {\it et al.}\cite{wu} have studied the collective 
modes, {\it i.e.} zeros of Re${\mathcal E}(q,\omega)$. 
The absolute value of $\chi(q,\omega)$ falls off as a 
function of frequency on a scale given by the width of the 
Landau levels, $\Gamma_0/\hbar=\sqrt{2\omega_c/\pi\tau}$. 
For half filled Landau levels $\Gamma_0/\hbar$ is the 
frequency range over which the 2DEG can respond to an external 
perturbation. As the polarization decreases with $\omega$ 
the effective interaction gets
enhanced, and competition between these two effects leads to a 
non-trivial temperature dependence in the same manner that led to 
a non-trivial dependence on the magnetic field. 

\begin{figure}
\epsfxsize=12cm
\epsfbox{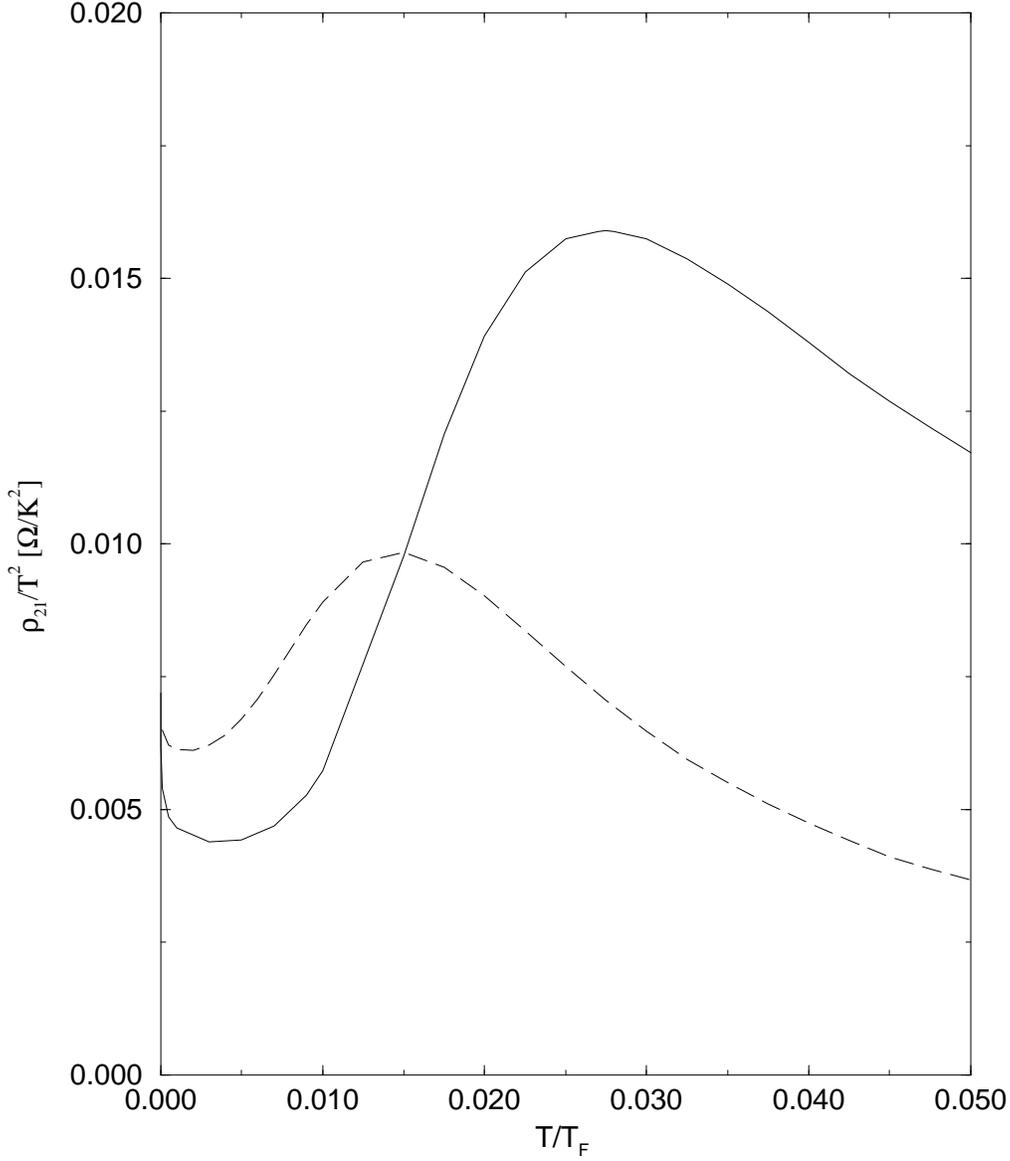}
\vspace{0.5cm}
\caption{
The scaled transresistivity $\rho_{21}/T^2$ as a function of temperature 
for filling factors $\nu=3$ (solid line) and $\nu=5$ (dashed line). 
Other parameters are as in Fig.~\ref{rhoofnu}. 
$\rho_{21}/T^2$ which in zero magnetic field is expected to be a 
constant, shows a maximum as a function of $T$ in intermediate magnetic 
fields. The enhancement arises at a temperature which is related to the 
width of the Landau levels as explained in the main text. 
}
\label{scrho1}
\end{figure}

In Fig.\ \ref{scrho1} we show plots of $\rho_{21}/T^2$ as a function 
of $T$ for $\nu=3$ and $\nu=5$. In contrast to the zero magnetic 
field case, $\rho_{21}/T^2$ shows a maximum at a peak temperature 
$T_{{\mathrm peak}}$.\cite{phonon1} 
The peak temperature is highest for the 
highest magnetic field (smallest $\nu$). 
The maximum in $\rho_{21}/T^2$ can be associated with excitations 
of states where the effective interaction is strong, {\it i.e.} 
where $\chi(q,\omega)$ is small. As pointed out above, $\chi(q,\omega)$ 
falls off over a frequency scale proportional to $\sqrt{B}$ 
which explains why $T_{{\mathrm peak}}(\nu=3)>T_{{\mathrm peak}}(\nu=5)$.
This prediction is in agreement with measurements of 
Rubel {\it et al.}\cite{rubel}
If $\rho_{21}/T^2$ were calculated using the static version of the 
screening function, it would be a monotonically decreasing function 
of temperature as shown in Ref.~\onlinecite{prl}.

\begin{figure}
\epsfxsize=12cm
\epsfbox{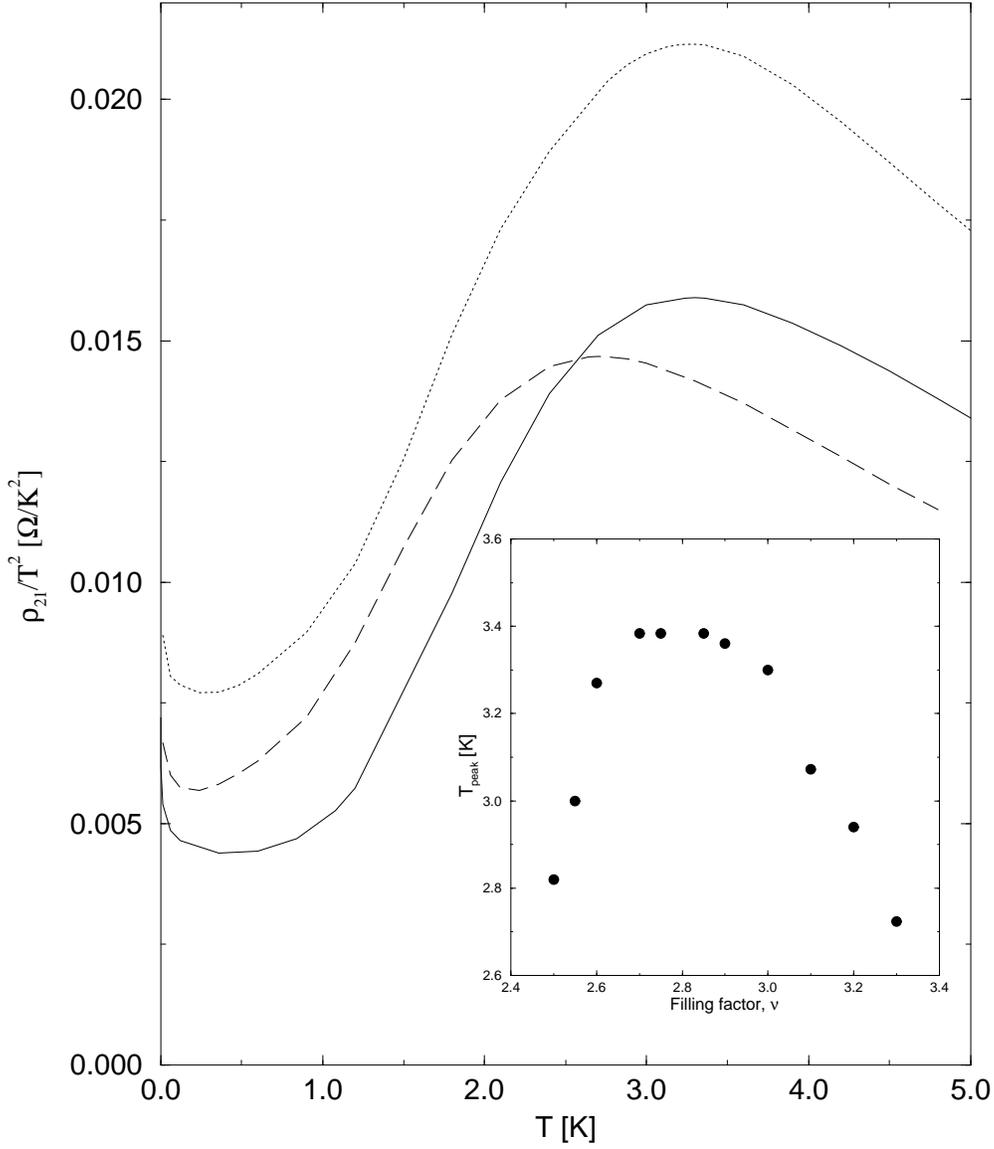}
\vspace{0.5cm}
\caption{
The scaled transresistivity $\rho_{21}/T^2$ as a function of $T$ for 
three filling factors, $\nu=2.6$ (dotted line), $\nu=3.0$ (solid line), 
$\nu=3.3$ (dashed line). Other parameters are as in Fig.~\ref{rhoofnu}. 
The temperature $T_{\rm peak}$ at which $\rho_{21}/T^2$ has a maximum 
depends on $\nu$.
$T_{\rm peak}$ as a function of $\nu$ is plotted in the inset. There is 
a maximum around $\nu=2.8$, {\it i.e.} just below an odd filling factor.
}
\label{scrho2}
\end{figure}

Having looked at the peak temperature for different odd filling factors, 
we now consider small changes of the filling factor around a given odd 
value.  Specifically, we examine 
$\nu=3\pm\delta\nu$. As the filling factor moves slightly away from an odd 
value, the system becomes less susceptible to perturbations; the 
polarization function falls off with frequency over a smaller scale. 
As a consequence, the screening function has a minimum at smaller 
frequencies which in turn implies that one
would expect  the peak temperature to become
smaller. In Fig.~\ref{scrho2} we plot the transresistivity as a function of
temperature for three different filling factors. The insert shows the 
peak temperature as a function of filling factor. We find that 
$T_{\rm peak}$ indeed has a (broad) maximum around $\nu=2.8$. 
The deviation of 0.2 away 
from $\nu=3$ is due to the general trend that $T_{\rm peak}$ 
increases with magnetic field (cf. previous discussion).

\section{Conclusion}\label{conclu}
To lowest order in the screened interlayer interaction,
the transconductivity of a pair of coupled two-dimensional electron gases 
is expressible in terms of three-body correlation functions, 
${\mathbf \Delta}$, called triangle functions, which 
depend on the microscopic details of each system. 
In this paper we have shown that for an isotropic system of 
non-interacting electrons scattering against random,
short range impurities, the triangle function is 
proportional to the imaginary part of the polarization 
function (Eq.~(\ref{mainres})) in the limit $\omega_c\tau\gg 1$. 
In this limit, we find that the transresistivity tensor is diagonal.
Including bandstructure effects, 
sufficiently energy-dependent intralayer scattering 
time,\cite{plasmonlong,halldrag} or correlations 
between the layers (in addition to the drag force)
may introduce a nondiagonal elements to the transresistivity tensor
({\it i.e.}, Hall component to the drag).

By numerical evaluations we have illustrated how the 
interplay between the screened interlayer {\it e--e} 
interaction and the phase-space available for scattering 
leads to non-trivial behavior of the transresistivity 
as a function of both magnetic field, where the characteristic 
is the twin-peak structure, and temperature dependence,
where $\rho_{21}/T^2$ should have a maximum at a temperature 
related to the width of the Landau levels. 

The results presented above are based on a relatively simple 
model for the polarization and screening functions.
We argue that this model is applicable as long as an independent 
electron picture describes the individual layers, and should 
in that regime give qualitatively correct results when the 
Landau levels are fully resolved. 

Whereas the specific models for the polarization and screening
functions break down at higher magnetic fields and/or lower 
temperatures, the general 
expression for the transconductivity in terms of the 
triangle functions remains valid (in the absence of interlayer
correlations) and is open to improvements.

\acknowledgements
We are grateful for rewarding discussions with Allan H. MacDonald. 
We also wish to thank Nick Hill, Holger Rubel and Tom Gramila
for detailed discussions of their experiments. 
This work was in part supported by the National Science Foundation 
under grant DMR-9416906. 
MCB is supported by the Danish Research Academy.

\appendix

\section{Properties of \protect${\bf\Delta}$}\label{app:deltaprop}

Since $\sigma_{21}(\omega=0)$ must be real, ${\bf\Delta}$ for each layer
must be purely real.  (To show this, assume that ${\bf\Delta}_1$ has an
imaginary component at ${\bf q}_0$ and $\omega_0$.  Then, for
a purely real ${\bf\Delta}_2$ and  $|V_{12}({\bf q},\omega)|^2
= C\,\delta({\bf q}-{\bf q}_0)\,\delta(\omega-\omega_0)$ 
$\sigma_{21}$ is not purely real, leading to a contradiction.)
Furthermore, ${\bf\Delta}$ for each layer must be gauge invariant, 
since all operators which make up $\Delta$ are gauge invariant.

Below, we give four symmetry properties of ${\bf\Delta}$. 
By definition, 
${\bf\Delta}({\bf x},\tau; {\bf x}',\tau'; {\bf x}'',\tau'')
= 
{\bf\Delta}({\bf x},\tau; {\bf x}'',\tau'; {\bf x}',\tau'')
$
which immediately implies from Eq.\ (\ref{ft_delta}), 
\begin{equation}
{\bf\Delta}({\bf q},{\bf q};\omega+i\eta,\omega-i\eta;{\bf B})
\equiv 
{\bf\Delta}(-{\bf q},-{\bf q};-\omega+i\eta,-\omega-i\eta;{\bf B}).
\end{equation}
Since ${\bf \Delta}$ is a vector quantity, in an isotropic system 
it must have the form ${\bf\Delta}({\bf q},{\bf q};\omega+i\eta,\omega-i\eta) 
= \Delta_{\|}(q,\omega+i\eta,\omega-i\eta,B) \hat{\bf q} + 
\Delta_\perp(q,\omega+i\eta,\omega-i\eta,B) (\hat{\bf q} \times \hat{\bf B})$.
{}From this, one can glean 
\begin{equation}
{\bf\Delta}({\bf q},{\bf q};\omega+i\eta,\omega-i\eta;{\bf B})
\equiv 
-{\bf\Delta}(-{\bf q},-{\bf q};\omega+i\eta,\omega-i\eta;{\bf B}).
\end{equation}
and
\begin{eqnarray}
{\Delta}_{\|} ({\bf q},{\bf q};\omega+i\eta,\omega-i\eta;{\bf B})
&=& {\Delta}_{\|}
({\bf q},{\bf q};\omega+i\eta,\omega-i\eta;-{\bf B}),
\nonumber\\
{\Delta}_\perp({\bf q},{\bf q};\omega+i\eta,\omega-i\eta;{\bf B})
&=& -\Delta_\perp({\bf q},{\bf q};\omega+i\eta,\omega-i\eta;-{\bf B}).
\end{eqnarray}
Finally, the Onsager relationship $\sigma_{21}^{\alpha\beta}({\bf B})
= \sigma_{12}^{\beta\alpha}(-{\bf B})$ implies 
\begin{equation}
{\bf\Delta}({\bf q},{\bf q};\omega+i\eta,\omega-i\eta;{\bf B})
\equiv 
{\bf\Delta}(-{\bf q},-{\bf q};-\omega-i\eta,-\omega+i\eta;-{\bf B}).
\end{equation}
Using the above four relationships, we know how any inversion
$\pm {\bf q}$, $\pm\omega$, $\pm \eta$ and $\pm {\bf B}$ 
affects ${\bf\Delta}$.

\section{Charge-vertex functions}\label{appvertex}
We first provide an approximation for the charge-vertex function 
which will also be useful for the purpose of later numerical 
evaluations. 

In the self consistent Born approximation which we have
adopted for the self-energy, the Landau level indices are not 
mixed, {\it i.e.} the Green functions remain diagonal (this
is equivalent to saying that the Landau levels are 
clearly resolved). Consistent with this, we can neglect 
coupling between Landau levels in the vertex functions, {\it i.e.} 
in the summation over $a$ and $b$ in Eq.~(\ref{vertex}) we set
$f_{b,m}({\mathbf q})=f_{b,m}({\mathbf q})\delta_{b,m}$
and $f_{n,a}(-{\mathbf q})=f_{n,a}(-{\mathbf q})\delta_{n,a}$.
For short range scatterers 
$U(k)$ is a constant and can be taken out of the integral. 
Then
\begin{eqnarray}\label{gamgam}
\Gamma({\mathbf q},n,m,i\epsilon+i\omega,i\epsilon)
&=&
f_{n,m}({\mathbf q}) \nonumber\\
&+&
(\Gamma_0/2\hbar)^2 {\mathcal I}(q,n,m)
{\mathcal G}(n,i\epsilon+i\omega){\mathcal G}(m,i\epsilon)
\Gamma({\mathbf q},n,m,i\epsilon+i\omega,i\epsilon) 
\end{eqnarray}
with
\begin{equation}\label{iii}
{\mathcal I}(q,n,m)=(-1)^{n+m}e^{-(\ell q)^2/2}
L_{n}^{m-n}((\ell q)^2/2)L_{m}^{n-m}((\ell q)^2/2).
\end{equation}

\begin{figure}
\epsfxsize=13.5cm
\epsfbox{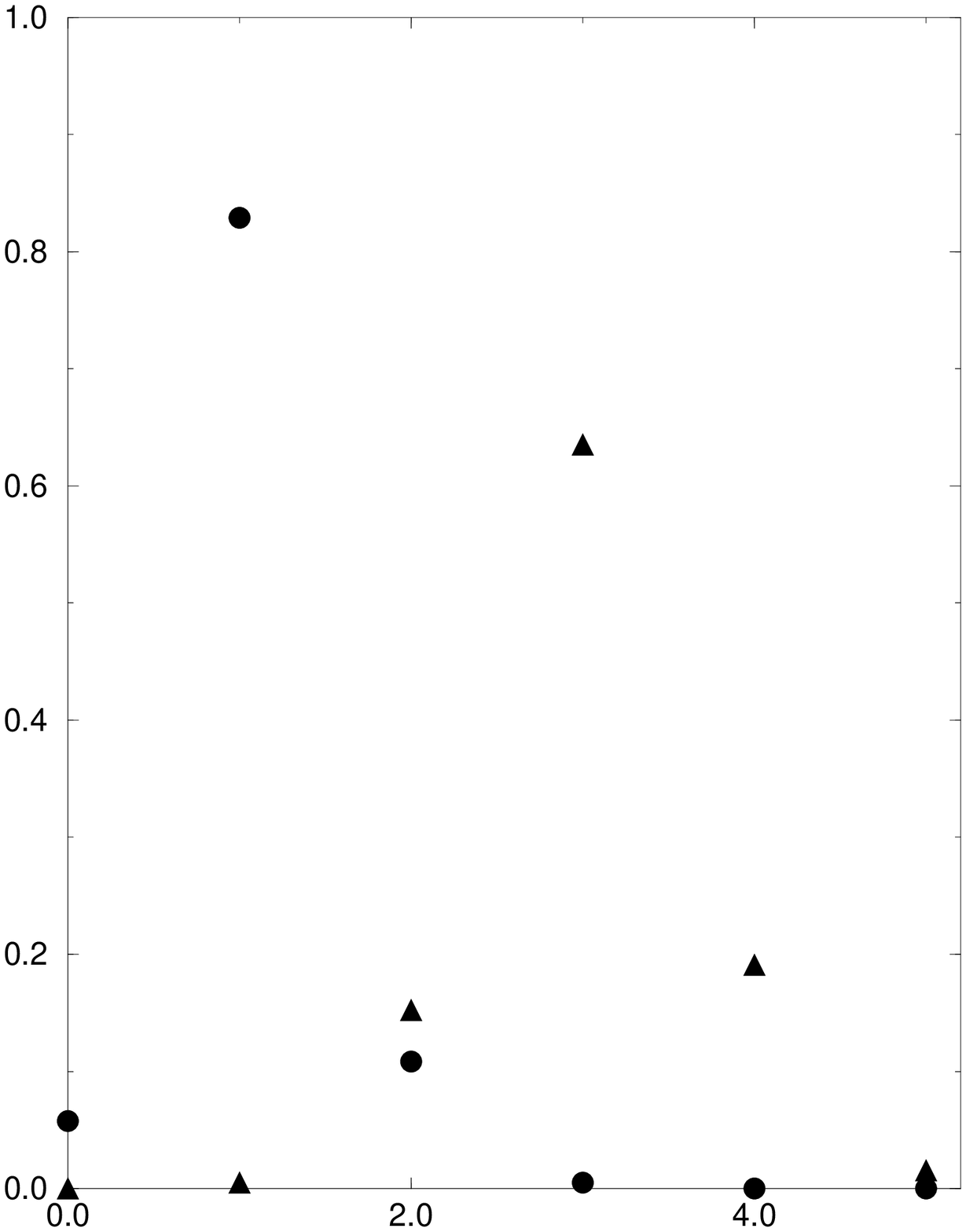}
\vspace{0.5cm}
\caption{
The function ${\mathcal I}(q,n,m)$ as a function of $m$ for $n=1$ (circles), 
and $n=3$ (triangles). In both cases $\ell q=0.35$.
}
\label{ii}
\end{figure}

The vertex function is a sum of a bare vertex and a correction;
we write Eq.~(\ref{gamgam}) as 
\begin{equation}
\Gamma^{\pm\pm}({\mathbf q},n,m,\epsilon_1,\epsilon_2)
=f_{n,m}({\mathbf q})+\delta^{\pm\pm}({\mathbf q},n,m,\epsilon_1,\epsilon_2)
\end{equation}
which defines $\delta^{\pm\pm}({\mathbf q},n,m,\epsilon_1,\epsilon_2)$.
We now show that the correction, $\delta$, is small as compared to 
the bare vertex unless $n=m$. 
{}From Eq.~(\ref{gamgam}) we see that this amounts to showing that
\begin{equation}\label{ggi}
(\Gamma_0/2\hbar)^2
G^{r}(n,\epsilon)G^{a}(m,\epsilon+\omega){\mathcal I}(q,n,m)\ll 1
\ \ \ \ \ ,\ n\not= m.
\end{equation}
We have chosen to consider $\Gamma^{+-}({\mathbf q},n,m,
\epsilon,\epsilon+\omega)$ as an (important) example. 
For $\epsilon\sim\varepsilon_n=(n+1/2)\omega_c$ and for 
$|\omega|\ll\omega_c$ we can approximate (from Eq.~(\ref{appp}))
\begin{equation}
G^{r}(n,\epsilon)G^{a}(m,\epsilon+\omega)\lesssim
\frac{2\hbar}{\Gamma_0}\frac{1}{(n-m)\omega_c},
\end{equation}
where we have used that $|G(n,\epsilon)|$ is of the order 
$2\hbar/\Gamma_0$ at its maximum.
We thus have to verify 
\begin{equation}
\frac{\Gamma_0}{2\hbar\omega_c}\frac{1}{(n-m)}{\mathcal I}
(q,n,m)\ll 1.
\end{equation}
In Fig.~\ref{ii} we plot the function ${\mathcal I}(q,n,m)$ 
as a function of $m$ for typical $q$ and $n$, and conclude
that the correction to the bare vertex is only appreciable 
when the Landau level indices of the two incoming Green functions are
equal. 
When $n=m$ in Eq.~(\ref{gamgam}), on the other hand, the correction 
$\delta$ is crucial for small frequencies, $\omega<1/\tau$. In this 
regime the correction to the bare vertex is responsible for the 
diffusive behavior which leads to a unique temperature dependence
of the transresistivity as we discuss in Sect.\ref{tempdep}. 

We now proceed to explain why (\ref{ggi}) makes the approximation
in Eq.~(\ref{approx}) valid.
{}From the left hand side of Eq.~(\ref{approx}) we have terms of 
the form (after using Eq.~(\ref{aprd}))
\begin{equation}\label{start}
\Gamma^{+-}({\mathbf q},n,m,\epsilon,\epsilon+\omega)
\Gamma^{--}(-{\mathbf q},m,n-1,\epsilon+\omega,\epsilon)
G^{r}(n,\epsilon)G^{a}(m,\epsilon+\omega).
\end{equation}
Since the correction, $\delta$, to the bare vertex 
is only appreciable when the Landau level indices, $n$ and $m$, are equal, 
we only have to keep terms from (\ref{start}) with at most one correction. 
There are two terms with exactly one correction: 
\begin{equation}
f_{n,m}({\mathbf q})\delta^{--}(-{\mathbf q},m,n-1,\epsilon+\omega,
\epsilon)G^{r}(n,\epsilon)G^{a}(m,\epsilon+\omega)
\end{equation}
and 
\begin{equation}
f_{m,n-1}(-{\mathbf q})\delta^{+-}({\mathbf q},n,m,\epsilon,
\epsilon+\omega)G^{r}(n,\epsilon)G^{a}(m,\epsilon+\omega).
\end{equation}
We now argue that the first of these can be neglected compared 
to the second.
The two terms should be summed over $n$ and $m$ (see Eq.~(\ref{fourp})).
Hence, we should compare
\begin{equation}\label{first}
\sum_{n,m}\sqrt{n+1}f_{n+1,m}({\mathbf q})\delta^{--}(-{\mathbf q},m,n,
\epsilon+\omega,\epsilon)G^{r}(n+1,\epsilon)G^{a}(m,\epsilon+\omega)
\end{equation}
and 
\begin{equation}
\sum_{n,m}\sqrt{n}f_{m,n-1}(-{\mathbf q})\delta^{+-}({\mathbf q},n,m,
\epsilon,\epsilon+\omega)G^{r}(n,\epsilon)G^{a}(m,\epsilon+\omega),
\end{equation}
where we have shifted the sum over $n$ in (\ref{first}) in order to 
make the corrections, $\delta$, directly comparable in order of magnitude. 
Since the product of Green functions is small when the Landau level indices are 
not equal, it is clear that the first term can be neglected when the 
sum over $n$ and $m$ is carried out. Hence the term in Eq.~(\ref{start}) 
is approximately
\begin{equation}
\Gamma^{+-}({\mathbf q},n,m,\epsilon,\epsilon+\omega)
f_{m,n-1}(-{\bf q})G^{r}(n,\epsilon)G^{a}(m,\epsilon+\omega).
\end{equation}
Other terms work out similarly.



\newpage

\newpage

\begin{thebibliography}{10}
\bibitem{phonon1} The phonon contribution, as discussed in 
Ref.~\onlinecite{phonon}, also contributes a peak in $\rho/T^2$ 
at approximately
the same temperatures at $B=0$.  Similar behavior may occur
at finite $B$.  However, a detailed calculation is needed in order to
determine the relative contributions of the Coulomb versus the phonon 
mediated interaction at $B\ne 0$.
\end{thebibliography}
\end{document}